\documentclass[8.5pt,twoside,twocolumn]{article}
\usepackage[super,sort&compress,comma]{natbib} 
\usepackage[version=3]{mhchem}
\usepackage{times,mathptmx}
\usepackage{sectsty}
\usepackage{balance} 

\usepackage{graphicx} 
\usepackage{lastpage}
\usepackage[format=plain,justification=raggedright,singlelinecheck=false,font=small,labelfont=bf,labelsep=space]{caption} 
\usepackage{fancyhdr}
\begin{document}

\newcommand{\commento}{}
\thispagestyle{plain}
\fancypagestyle{plain}{
\renewcommand{\headrulewidth}{1pt}}
\renewcommand{\thefootnote}{\fnsymbol{footnote}}
\renewcommand\footnoterule{\vspace*{1pt}%
\hrule width 3.4in height 0.4pt \vspace*{5pt}} 
\setcounter{secnumdepth}{5}

\makeatletter 
\def\subsubsection{\@startsection{subsubsection}{3}{10pt}{-1.25ex plus -1ex minus -.1ex}{0ex plus 0ex}{\normalsize\bf}} 
\def\paragraph{\@startsection{paragraph}{4}{10pt}{-1.25ex plus -1ex minus -.1ex}{0ex plus 0ex}{\normalsize\textit}} 
\renewcommand\@biblabel[1]{#1}            
\renewcommand\@makefntext[1]%
{\noindent\makebox[0pt][r]{\@thefnmark\,}#1}
\makeatother 
\renewcommand{\figurename}{\small{Fig.}~}
\sectionfont{\large}
\subsectionfont{\normalsize} 

\fancyfoot{}
\fancyfoot[RO]{\footnotesize{\sffamily{1--\pageref{LastPage} ~\textbar  \hspace{2pt}\thepage}}}
\fancyfoot[LE]{\footnotesize{\sffamily{\thepage~\textbar\hspace{3.45cm} 1--\pageref{LastPage}}}}
\fancyhead{}
\renewcommand{\headrulewidth}{1pt} 
\renewcommand{\footrulewidth}{1pt}
\setlength{\arrayrulewidth}{1pt}S
\setlength{\columnsep}{6.5mm}
\setlength\bibsep{1pt}

\twocolumn[
\begin{@twocolumnfalse}
\noindent\LARGE{\textbf{Sorting of Chiral Microswimmers}}
\vspace{0.6cm}

\noindent\large{\textbf{Mite Mijalkov \textit{$^{a}$} and Giovanni Volpe,$^{\ast}$\textit{$^{a}$}}}\vspace{0.5cm}

\noindent\textit{\small{\textbf{Received Xth XXXXXXXXXX 20XX, Accepted Xth XXXXXXXXX 20XX\newline
First published on the web Xth XXXXXXXXXX 200X}}}

\noindent \textbf{\small{DOI: 10.1039/b000000x}}
\vspace{0.6cm}

\noindent \normalsize{
Microscopic swimmers, e.g., chemotactic bacteria and cells, are capable of directed motion by exerting a force on their environment. For asymmetric microswimmers, e.g., bacteria, spermatozoa and many artificial active colloidal particles, a torque is also present leading in two dimensions to circular motion and in three dimensions to helicoidal motion with a well-defined chirality. Here, we demonstrate with numerical simulations in two dimensions how the chirality of circular motion couples to chiral features present in the microswimmer environment. Levogyre and dextrogyre microswimmers as small as $50\,\mathrm{nm}$ can be separated and selectively trapped in \emph{chiral flowers} of ellipses. Patterned microchannels can be used as \emph{funnels} to rectify the microswimmer motion, as \emph{sorters} to separate microswimmers based on their linear and angular velocities, and as \emph{sieves} to trap microswimmers with specific parameters. We also demonstrate that these results can be extended to helicoidal motion in three dimensions.
}
\vspace{0.5cm}

\end{@twocolumnfalse}
]

\section{Introduction}

\footnotetext{\textit{$^{a}$~Department of Physics, Bilkent University, Cankaya, Ankara 06800, Turkey.; E-mail: giovanni.volpe@fen.bilkent.edu.tr}}

Differently from simple Brownian particles, whose motion is dominated by random thermal fluctuations, microswimmers are capable of directed motion. Many microorganisms can actively explore their environment, e.g., \emph{E. coli} bacteria and white blood cells \cite{Berg2004}. Recently, several artificial microswimmers have also been demonstrated, e.g., self-diffusiophoretic Janus microparticles \cite{Ismagilov2002,Paxton2005,Vicario2005,Golestanian2005,Howse2007,Pantarotto2008,Tierno2008,Ghosh2009,Snezhko2009,Ebbens2010,Volpe2011,Buttinoni2012}. These microswimmers hold the promise of major applications in several fields, e.g., to deliver drugs within tissues, to localize pollutants in soils and to perform tasks in lab-on-a-chip devices \cite{Weibel2005,Ford2007,Chin2007,Yang2012}.

\begin{figure}[h]
\centering
\includegraphics[width=8.5cm]{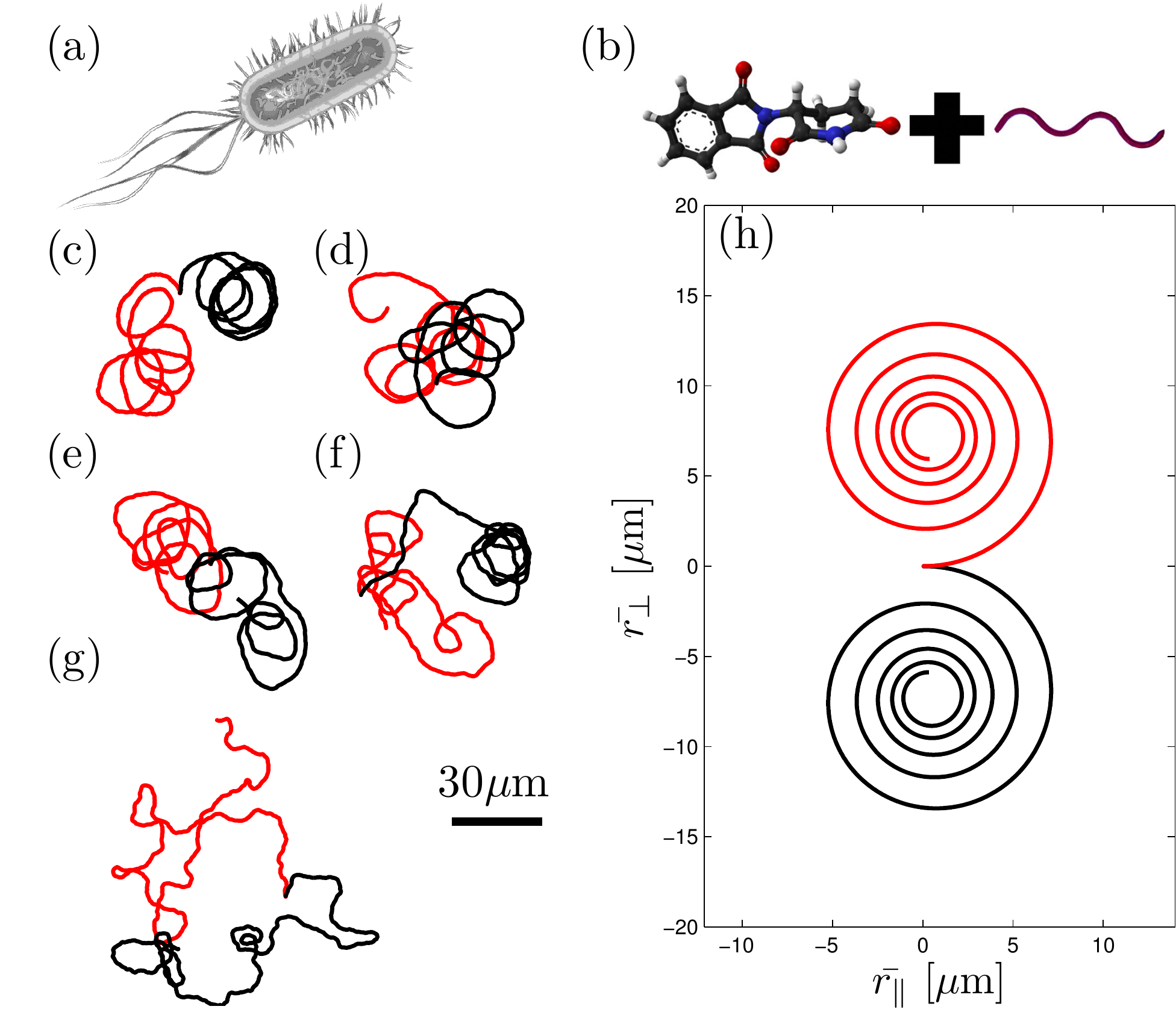}
\caption{Chiral microswimmers. $(a)$ \emph{Escherichia coli} bacteria perform a characteristic chiral motion in the proximity of a surface. $(b)$ Active chiral molecules can be obtained by chemically attaching a chiral molecule with a chiral propeller, e.g., a flagellum. $(c-g)$ The trajectories of chiral levogyre (red) and dextrogyre (black) microswimmers with different radii ($R = 1000$, $500$, $250$, $125$, and $50\,\mathrm{nm}$ for $(c)$, $(d)$, $(e)$, $(f)$ and $(g)$ respectively, see Tab.~\ref{tbl:table} for the other parameters) are qualitatively similar as long as the P\'eclet number is kept constant and the time is scaled accordingly ($t=10\,\mathrm{s}$, $2.5\,\mathrm{s}$, $625\,\mathrm{ms}$, $157\,\mathrm{ms}$ and $25\,\mathrm{ms}$ for $(c)$, $(d)$, $(e)$, $(f)$ and $(g)$ respectively). See also the supplementary movies 1, 2 and 3 corresponding to $(c)$, $(d)$ and $(e)$ respectively. $(h)$ Average of $10^5$ trajectories starting at $[x(0),\, y(0)] = [0\, \mathrm{\mu m},\, 0\, \mathrm{\mu m}]$ with $\varphi(0) = 0$ for levogyre (red) and dextrogyre (black) microswimmers as in $(c)$.
}
\label{fgr:figure1}
\end{figure}

In order to perform active motion, a microswimmer must exert a force on its surroundings \cite{Downton2009,Drescher2010}. If this driving force acts along the line of motion, the microswimmer will move along a straight line just perturbed by random Brownian fluctuations \cite{Howse2007}. However, often the microswimmer is asymmetric so that the driving force and the propulsion direction are not aligned \cite{Teeffelen2008}. This results in the microswimmer exerting a torque and in chiral active Brownian motion, i.e. in two dimensions circular motion and in three dimensions helicoidal motion with a well-defined chirality. There are many natural examples of chiral microswimmers. For example, \emph{E. coli} bacteria (Fig.~\ref{fgr:figure1}a) and spermatozoa undergo helicoidal motion, which becomes two-dimensional chiral active Brownian motion when moving near boundaries\cite{DiLuzio2005,Lauda2006,Friedrich2009,Lemelle2010,DiLeonardo2011,Su2012,Denissenko2012}. Also, artificial microswimmers show characteristic chiral trajectories when they are asymmetrical either by engineering or by chance. Moving down to the nanoscale, we can also consider molecular-sized chiral microswimmers, which can be obtained by joining a chiral molecule with a chiral propeller, e.g., a flagellum (Fig.~\ref{fgr:figure1}b).

Sorting microswimmers based on their swimming properties, e.g. velocity, angular velocity and chirality, is of utmost importance for various branches of science and engineering. Genetically engineered bacteria can be sorted based on phenotypic variations of their motion \cite{Berg2004}. Velocity-based spermatozoa selection can be employed to enhance the success probability in the artificial fertilization techniques \cite{Guzik2001}. Considering the intrinsic variability of microfabrication techniques, the efficiency of artificial microswimmers for a specific task, e.g., drug-delivery or bioremediation, can be increased by selecting only the ones with the most appropriate swimming properties. Finally, the separation of levogyre and dextrogyre chiral molecules can be more effectively accomplished by chemically coupling them to chiral propellers, sorting the resulting chiral microswimmers and finally detaching the propellers. This is important because, e.g., often only one specific chirality is needed by the chemical and pharmaceutical industry \cite{Ahuja2011} and can hardly be achieved by  mechanical means due to the extremely small Reynolds numbers \cite{Marcos2009}, 

In this article, we numerically demonstrate that chiral microswimmers can be sorted on the basis of their swimming properties by employing some simple static patterns in their environment. Even though we demonstrate most of the results using two-dimensional chiral microswimmers moving within two-dimensional patterned environments, we also show that these results can be adapted to the case of three-dimensional chiral microswimmers. We show that a \emph{chiral flower}, i.e. a chiral structure formed by some tilted ellipses arranged in a circle, can trap microswimmers with a specific chirality and can, therefore, be used to separate a racemic mixture. We also demonstrate that a patterned microchannel can be used  as a \emph{funnel} to rectify the motion of chiral microswimmers, as a \emph{sorter} of microswimmers based on their linear and angular velocities, and as a \emph{sieve} to trap microswimmers with specific parameters. All these phenomena can be scaled down to smaller microparticles as long as the P\'eclet number is maintained constant.

\section{Model}

\begin{table*}
\small
\caption{Microswimmer parameters used in the simulations. From the radius $R$ the rotational diffusion coefficient $D_R$ and the translational diffusion coefficient $D_T$ are obtained using Eqs.~(\ref{eqn:DR}) and (\ref{eqn:DT}) respectively. The linear velocity $v$ and the angular velocity $\omega$ are rescaled in order to maintain the P\'eclet number $Pe = \frac{R v}{D_T}$ constant.}\label{tbl:table}
\begin{tabular*}{\textwidth}{@{\extracolsep{\fill}}llllll}
\hline
$R \, \mathrm{(nm)}$ 
& $D_R \, \mathrm{(rad^2 \, s^{-1})}$ 
& $D_T \, \mathrm{(\mu m^2 \, s^{-1})}$
& $v \, \mathrm{(\mu m \, s^{-1})}$ 
& $\Omega \, \mathrm{(rad \, s^{-1})}$ 
& $P_e$ 
\\
\hline
1000 & 0.16 & 0.22 & 3.13e+1 & $\pm 3.14$ & 142 \\
500 & 1.32 & 0.44 & 1.25e+2 & $\pm 1.26$e+1 & 142 \\
250 & 10.54 & 0.88 & 5.00e+2 & $\pm 5.03$e+1 & 142 \\
125 & 84.4 & 1.76 & 2.00e+3 & $\pm 2.00$e+2 & 142 \\
50 & 1320 & 4.4 & 1.25e+4 & $\pm 1.25$e+3 & 142 \\
\hline
\end{tabular*}
\end{table*}

The motion of chiral microswimmers arises from the combined actions of random diffusion, an internal self-propelling force and a torque. We describe such motion using a model based on the one proposed in Ref. \citenum{Teeffelen2008} to describe two-dimensional chiral microswimmers performing circular active Brownian motion. The position $[x(t),\, y(t)]$ of a spherical particle with radius $R$ undergoes Brownian diffusion with translational diffusion coefficient
\begin{equation}\label{eqn:DT}
D_T = \frac{k_B T}{6 \pi \eta R}
\end{equation} 
where $k_B$ is the Boltzmann constant, $T$ is the temperature and $\eta$ is the fluid viscosity. The particle self-propulsion results in a directed component of the motion, whose speed $v$ we will assume to be constant and whose direction depends on the particle orientation $\varphi(t)$. $\varphi(t)$ undergoes rotational diffusion with rotational diffusion coefficient 
\begin{equation}\label{eqn:DR}
D_R = \frac{k_B T}{8 \pi \eta R^3}.
\end{equation}
In chiral microswimmers, $\varphi(t)$ also rotates with angular velocity $\Omega$ as a consequence of the torque acting on the particle. The sign of $\Omega$ determines the microswimmer chirality. The resulting set of Langevin equations describing this motion is: 
\begin{equation}\label{eqn:langevin}
\left\{\begin{array}{ccc}
d \varphi(t) & = & \Omega \; dt + \sqrt{2 D_R} \; dW_{\varphi} \\
d x(t) & = & v \sin(\varphi(t)) \; dt + \sqrt{2 D_T} \; dW_x \\
d y(t) & = & v \cos(\varphi(t)) \; dt + \sqrt{2 D_T} \; dW_y
\end{array}\right.
\end{equation}
where $W_{\varphi}$, $W_x$ and $W_y$ are independent Wiener processes. Inertial effects are neglected because of the low Reynolds number regime. Equations~(\ref{eqn:langevin}) are then solved using some standard finite-difference numerical methods \cite{Kloeden1992,Volpe2013}. We have assumed that the microswimmers are spherical, e.g., active Janus microparticles \cite{Howse2007,Volpe2011}, but these equations can be straightforwardly generalized to elongated microswimmers, e.g., nanorods and bacteria, by employing diffusion matrices instead of diffusion constants \cite{Fernandes2002,Kummel2013}. For the three-dimensional chiral motion simulations, we employed a straightforward generalization of the previous model based on Ref. \citenum{Wittkowski2012}.

In patterned environments, the microswimmers will often encounter obstacles \cite{Volpe2011}. In these interactions, when a microswimmer gets in contact with an obstacle, it landslides along the obstacle until its motion orientation points away from the obstacle. Numerically, this is implemented by suppressing the motion component perpendicular to the obstacle perimeter when pointing onwards the obstacle. Even though this model does not include hydrodynamic effects between the microswimmers and the structures, it has been shown to describe accurately the interaction trajectories of microswimmers near an obstacle in Ref. \citenum{Volpe2011}.

\section{Homogeneous Environments}

The red line in Fig.~\ref{fgr:figure1}c is a sample trajectory for a levogyre microswimmer with $R = 1000 \, \mathrm{nm}$ (see Tab.~\ref{tbl:table} for the full list of parameters) moving in an environment free from obstacles. It bends counterclockwise tracking almost circular trajectories just disturbed by Brownian fluctuations. Changing the chirality sign to dextrogyre (black line in Fig.~\ref{fgr:figure1}c), the trajectory behaves similarly but bending clockwise. This characteristic behavior becomes clearer considering the ensemble average of many trajectories. In Fig.~\ref{fgr:figure1}h, we plot the average of $10^5$ trajectories starting at $[x(0),\, y(0)] = [0\, \mathrm{\mu m},\, 0\, \mathrm{\mu m}]$ with $\varphi(0) = 0$ with levogyre (red line) and dextrogyre (black line) chirality; the average trajectories describe a \emph{spira mirabilis} whose orientation depends on the motion chirality \cite{Teeffelen2008}. The dimensions of the \emph{spire mirabilis} define a length scale for the rotation of the swimmers, which is relevant in the presence of patterns in the environment.

Similar results are obtained for smaller particles as long as the P\'eclet number is kept constant. This is achieved by rescaling $t$, $v$ and $\Omega$ according to $R^{-2}$. For example, in Figs.~\ref{fgr:figure1}d-g, the trajectories for $R = 500$, $250$, $125$, and $ 50 \, \mathrm{nm}$ (see Tab.~\ref{tbl:table} for the full list of parameters). Even though some qualitative resemblance between these trajectories and the ones for $R = 1000 \, \mathrm{nm}$ (Fig.~\ref{fgr:figure1}c) can be spotted, as the particle size decreases the trajectories become less deterministic due to the fact that the rotational diffusion, responsible for the reorientation of the particle direction, scales according to $R^{-3}$ (Eq.~(\ref{eqn:DR})) while the translation diffusion only according to $R^{-1}$ (Eq.~(\ref{eqn:DT})).

\begin{figure}[h]
\centering
\includegraphics[width=8.5cm]{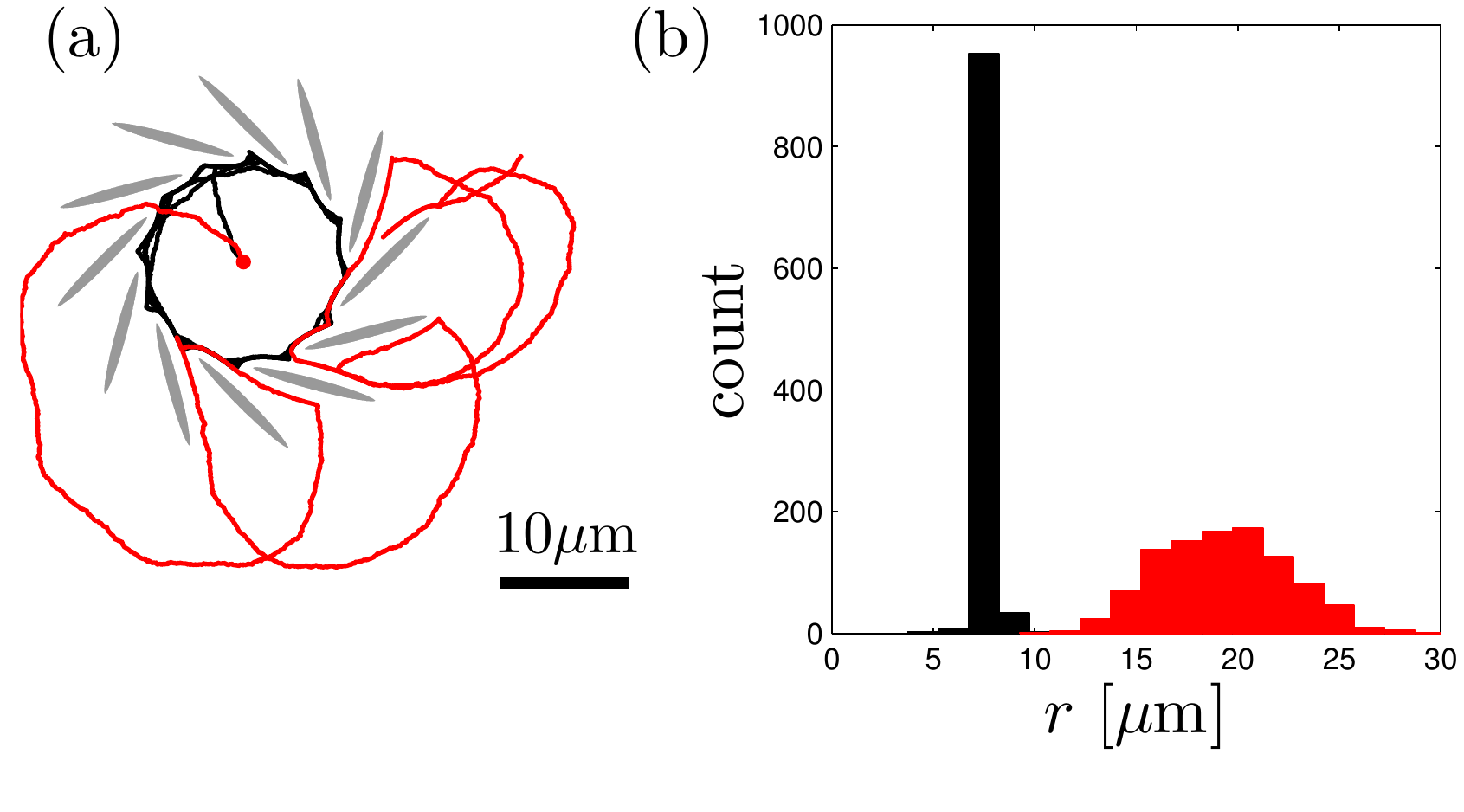}
\caption{Chiral microswimmers in a chiral environment. $(a)$ $10\,\mathrm{s}$ trajectories of a levogyre (red line) and dextrogyre (black line) microswimmer with $R=1000\,\mathrm{nm}$ (Tab.~\ref{tbl:table}) in a chiral flower of ellipses (shaded areas). See also the supplementary movie 4. $(b)$ Radial position distribution after $10\,\mathrm{s}$ of 1000 levogyre (red histogram) and dextrogyre (black histogram) microswimmers starting from the center of the chiral flower at $t=0\,\mathrm{s}$.
}
\label{fgr:figure2}
\end{figure}

\section{Patterned Environments}

\subsection{Chirality Separation} 

The microswimmers can be selected on the basis of the sign of their motion chirality in the presence of some chiral patterns in the environment, e.g., an arrangement of tilted ellipses (or any other elongated shape) along a circle forming what we call a \emph{chiral flower}. The shaded areas in Fig.~\ref{fgr:figure2}a depict a chiral flower where the ellipses are arranged along a circle with radius $11\,\mathrm{\mu m}$. The dimensions of this structure are chosen to be comparable with the characteristic length scale defined by the \emph{spire mirabilis} (\ref{fgr:figure1}). The levogyre microswimmer (red line in Fig.~\ref{fgr:figure2}a) is able to enter and exit the chiral flower without difficulty; however, a dextrogyre microswimmer (black line in Fig.~\ref{fgr:figure2}a) is trapped within the chiral flower. Fig.~\ref{fgr:figure2}b shows the radial position distribution of $10^5$ microswimmers after $10\,\mathrm{s}$ from when they are released from the center of the chiral flower: the levogyre microswimmers (red histogram) are outside the chiral flower, while the dextrogyre ones (black histogram) are trapped inside.

\begin{figure}[h]
\centering
\includegraphics[width=8.5cm]{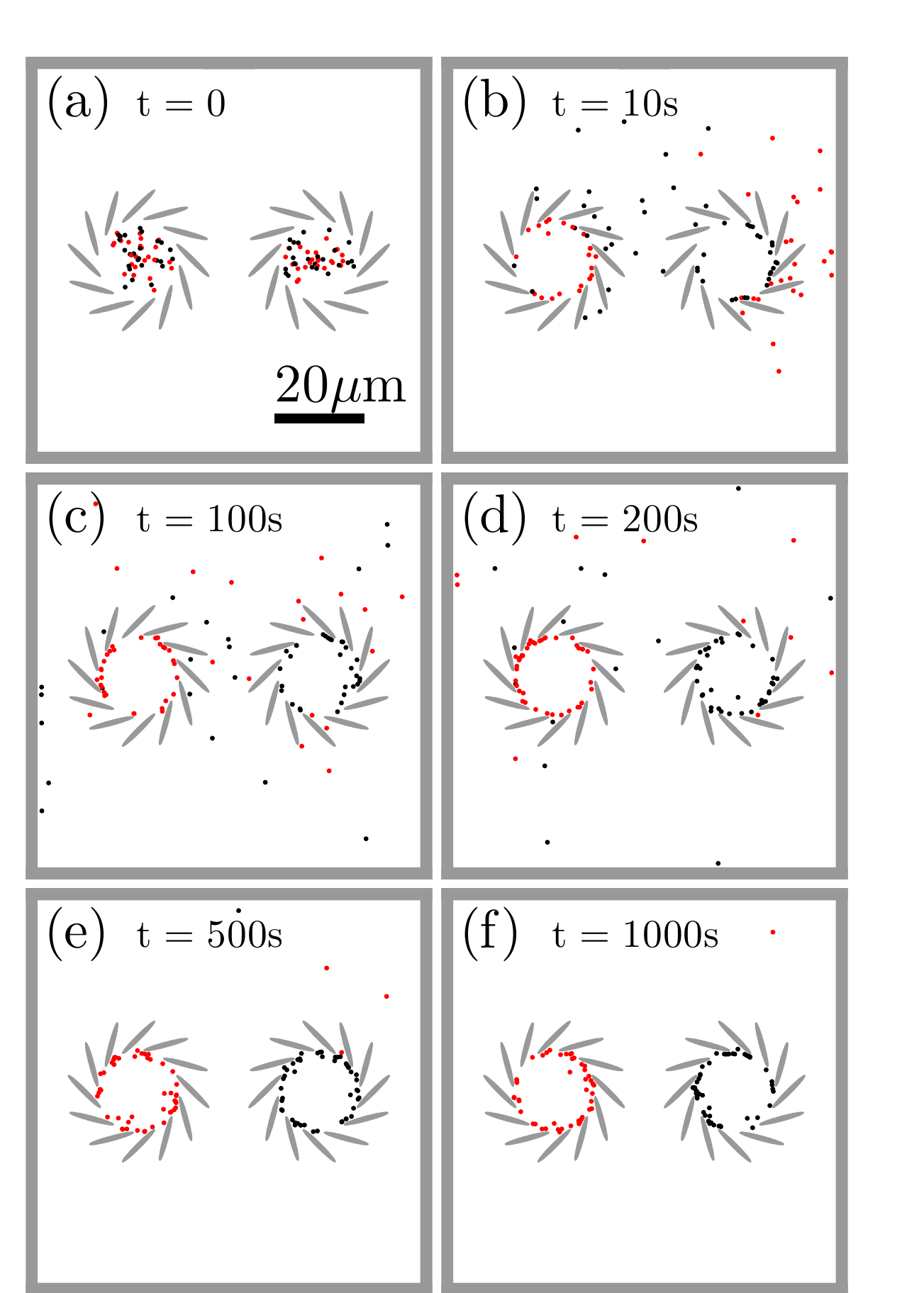}
\caption{Separation of levogyre and dextrogyre microswimmers. $(a)$ At $t=10\,\mathrm{s}$, the microswimmers ($R=1000\,\mathrm{nm}$, Tab.~\ref{tbl:table}) are released inside two chiral flowers with opposite chirality. $(b-f)$ As time progresses, the levogyre (black symbols) microswimmers are trapped in the right chiral flower and the dextrogyre (red symbols) ones in the left chiral flower. See also the supplementary movie 5.
}
\label{fgr:figure3}
\end{figure}

\begin{figure}[h]
\centering
\includegraphics[width=8.5cm]{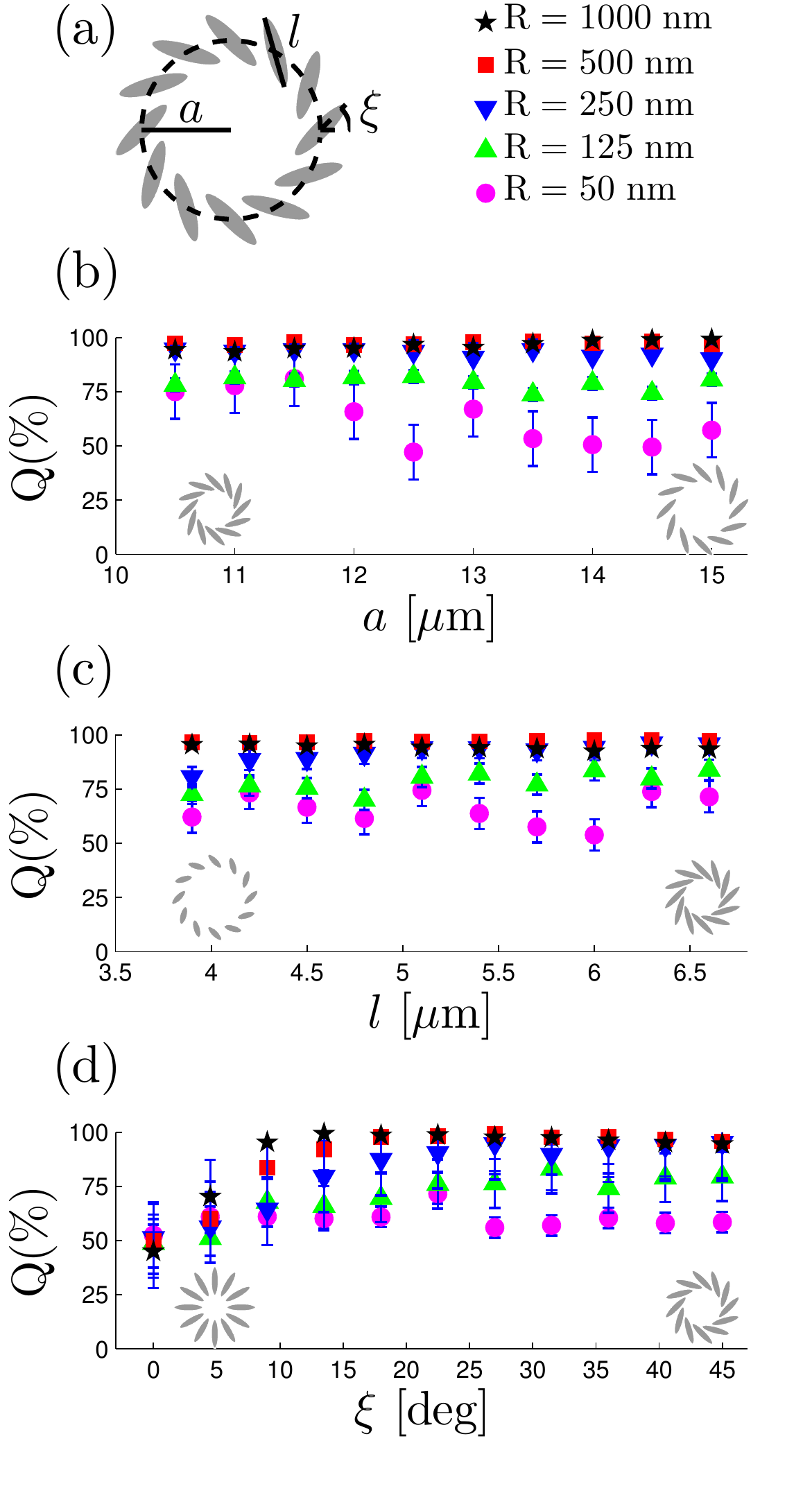}
\caption{Sorting efficiency. $(a)$ Parameters of the chiral flower: $a$ is the radius of the flower; $l$ is the length of the ellipses; and $\xi$ is the angle of the ellipses. (b-d) Sorting efficiency $Q$ (Eq.~(\ref{eq:q})) as a function of $(b)$ $a$, $(c)$ $l$, and $(d)$ $\xi$ for various particle sizes (Tab.~\ref{tbl:table}) using the configuration in Fig.~\ref{fgr:figure3}. Each datapoint is calculated using 50 levogyre and 50 dextrogyre particles placed inside the chiral flowers. The error bars represent one standard deviation repeating the numerical simulations 10 times.
}
\label{fgr:figure3n}
\end{figure}

We can use the previous observation to devise a simple device to separate and trap microswimmers with different chiralities. As shown in Fig.~\ref{fgr:figure3}, we use two chiral flowers with opposite chiralities enclosed in a box where the microswimmers can move freely. We start at time $t=0\,\mathrm{s}$ (Fig.~\ref{fgr:figure3}a) with a racemic mixture placed inside each of the chiral flowers. Already at $t=10\,\mathrm{s}$ (Fig.~\ref{fgr:figure3}b), most of the levogyre (dextrogyre) microswimmers have escaped the right (left) chiral flower, while the opposite chirality ones remain trapped. In the subsequent time, the free microswimmers explore the space of the box util they get trapped by the corresponding chiral flower (Fig.~\ref{fgr:figure3}c-d). Almost all microswimmers are stably trapped by $t=500\,\mathrm{s}$ (Fig.~\ref{fgr:figure3}e-f). In order to quantify the sorting efficiency, we introduce the parameter
\begin{equation}\label{eq:q}
Q = \frac{1}{2} \left[ \frac{L_L}{L_L+D_L} + \frac{D_D}{L_D+D_D} \right] \times 100\%,
\end{equation}
where $L_L$ ($D_L$) is the number of levogyre (dextrogyre) microswimmers in the levogyre flower and $D_D$ ($L_D$) is the number of dextrogyre (levogyre) microswimmers in the dextrogyre flower calculated after $500\,\mathrm{s}$. We easily obtain $Q \approx 100 \%$ for a wide range of parameters of the chiral flower, as we show in Fig.~\ref{fgr:figure3n}. The time to separate the microswimmers is not significantly affected by the parameters of the chiral flowers, but mainly by the size of the containing box. Furthermore, very high sorting efficiencies are obtained using the same chiral flowers also for smaller particles down to $50\,\mathrm{nm}$, despite the dominant randomness of their trajectories (Fig.~\ref{fgr:figure3n}).

\subsection{Sorting by Velocity}

\begin{figure}[h!]
\centering
\includegraphics[width=8.5cm]{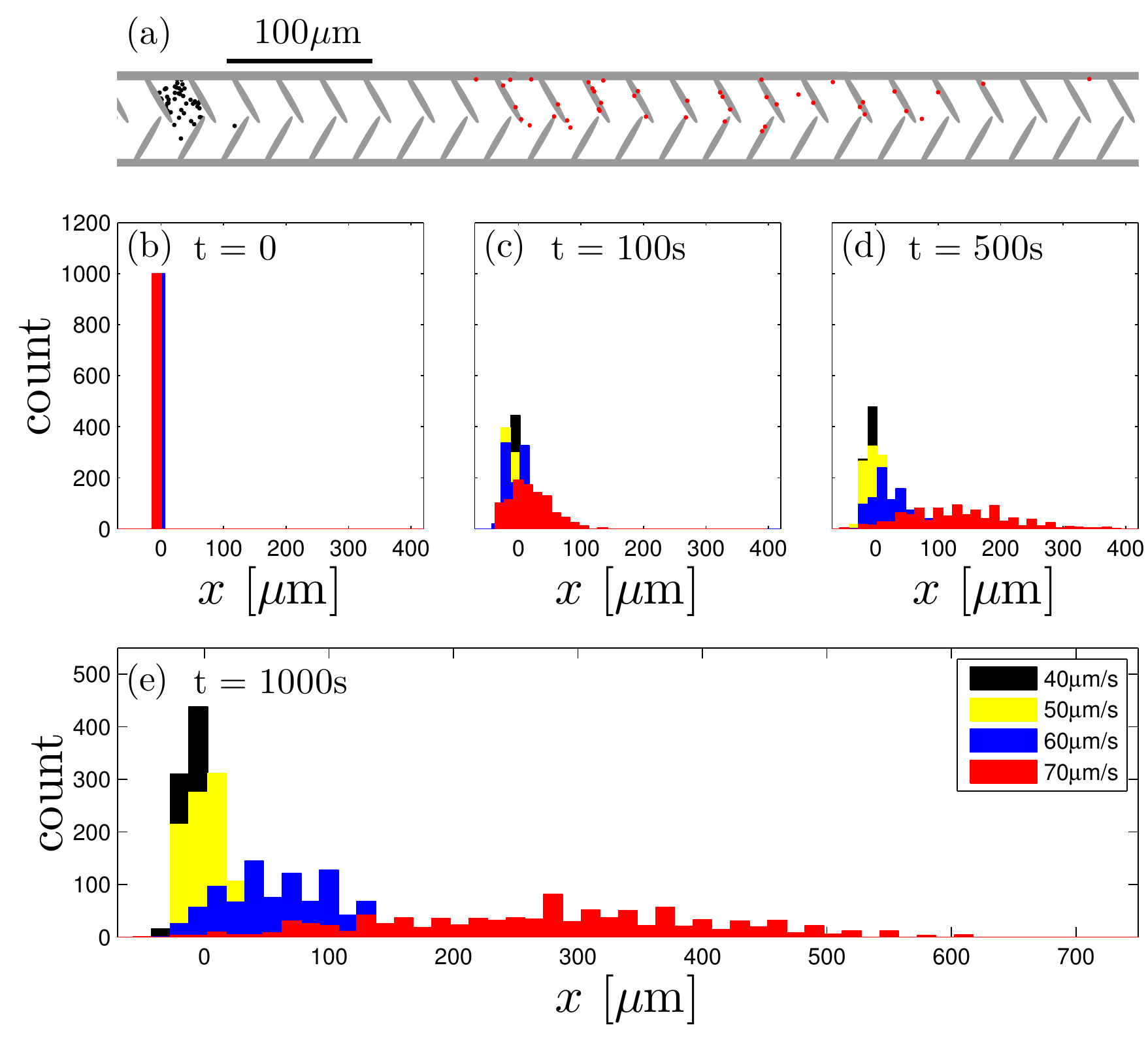}
\caption{Linear velocity based sorting in a microchannel. $(a)$ Patterned microchannel (grey areas) and position of levogyre $1000\,\mathrm{nm}$ microswimmers with $v=40$ (black symbols) and $70\,\mathrm{\mu m/s}$ (red symbols) $1000\,\mathrm{s}$ after they have been released from position $x = 0\,\mathrm{\mu m}$ (other parameters as in Tab.~\ref{tbl:table}). See also the supplementary movie 6. $(b-e)$ Histograms at various times of 1000 particles with $v =40,\,50,\,60$ and $70\, \mathrm{\mu m/s}$ released at $t=0\,\mathrm{s}$ from $x = 0\,\mathrm{\mu m}$.
}
\label{fgr:figure4}
\end{figure}

A patterned microchannel can be used to sort the microswimmers on the basis of their linear velocity. We show an example of such microchannel in Fig.~\ref{fgr:figure4}a (grey areas). It is formed by a series of elliptical structures inserted on the channel walls at an angle and slightly shifted between the top and bottom sides. In this way the channel is itself chiral.  Chiral microswimmers placed at position $x = 0\,\mathrm{\mu m}$ at $t=0\,\mathrm{s}$ behave differently depending on their linear velocity. For example, at $t=1000\,\mathrm{s}$, levogyre microswimmers with $v = 70\, \mathrm{\mu m/s}$ (red symbols in Fig.~\ref{fgr:figure4}a) have propagated towards the right several hundreds microns, while slower microswimmers with $v = 40\, \mathrm{\mu m/s}$ (black symbols in Fig.~\ref{fgr:figure4}a) are trapped near the initial position; both microswimmers have the same chirality and angular velocity $\Omega = +3.1\,\mathrm{rad/s}$. The patterned microchannel works as a \emph{funnel} to rectify the motion of chiral microswimmers, as demonstrated from the fact that the microswimmers move towards the right. The microswimmer motion in the channel can be thought of as a ratchet driven by the microswimmers own self-propulsion \cite{Reimann2002}. The microchannel works also as a \emph{sieve} to trap microswimmers with specific parameters, e.g., in Fig.~\ref{fgr:figure4} only microswimmers with $v>40\, \mathrm{\mu m/s}$ can propagate to the right. Changing the structure of the channel it is possible to change the velocity threshold below which the microswimmers start propagating. Finally, it is also possible to use the channel as a microswimmer \emph{sorter} based on their linear, as shown in Figs.~\ref{fgr:figure4}b-e. When we place at $t=0\,\mathrm{s}$ a mixture of particles with $v =40,\,50,\,60$ and $70\, \mathrm{\mu m/s}$ at position $x = 0\,\mathrm{\mu m/s}$ (Fig.~\ref{fgr:figure4}b), we observe that the probability distributions of particles with different speed separate over time (Figs.~\ref{fgr:figure4}c-e). Faster particles propagate further along the channel while the slower particles are held back. The separation efficiency increases with time as the particles propagate further along the channel.

\subsection{Sorting by Angular Velocity}

The patterned microchannel introduced in the previous subsection can also be used to sort particles according to their angular velocity. In Fig.~\ref{fgr:figure4}a, we show that microswimmers with $v =40\, \mathrm{\mu m/s}$ and $\Omega = 3.1\, \mathrm{rad/s}$ (black symbols) are trapped at their initial position after $10000\, \mathrm{s}$, while microswimmers with $\Omega = 2.2\, \mathrm{rad/s}$ (red symbols) can propagate. Over time, it is possible to separate particles with smaller differences in angular velocity, as shown by the histograms in Fig.~\ref{fgr:figure4}b-e for $\Omega =2.2,\,2.5,\,2.8$ and $3.1\, \mathrm{\mu m/s}$.

\subsection{Sorting of 3D Chiral Microswimmers}

\begin{figure}[h!]
\centering
\includegraphics[width=8.5cm]{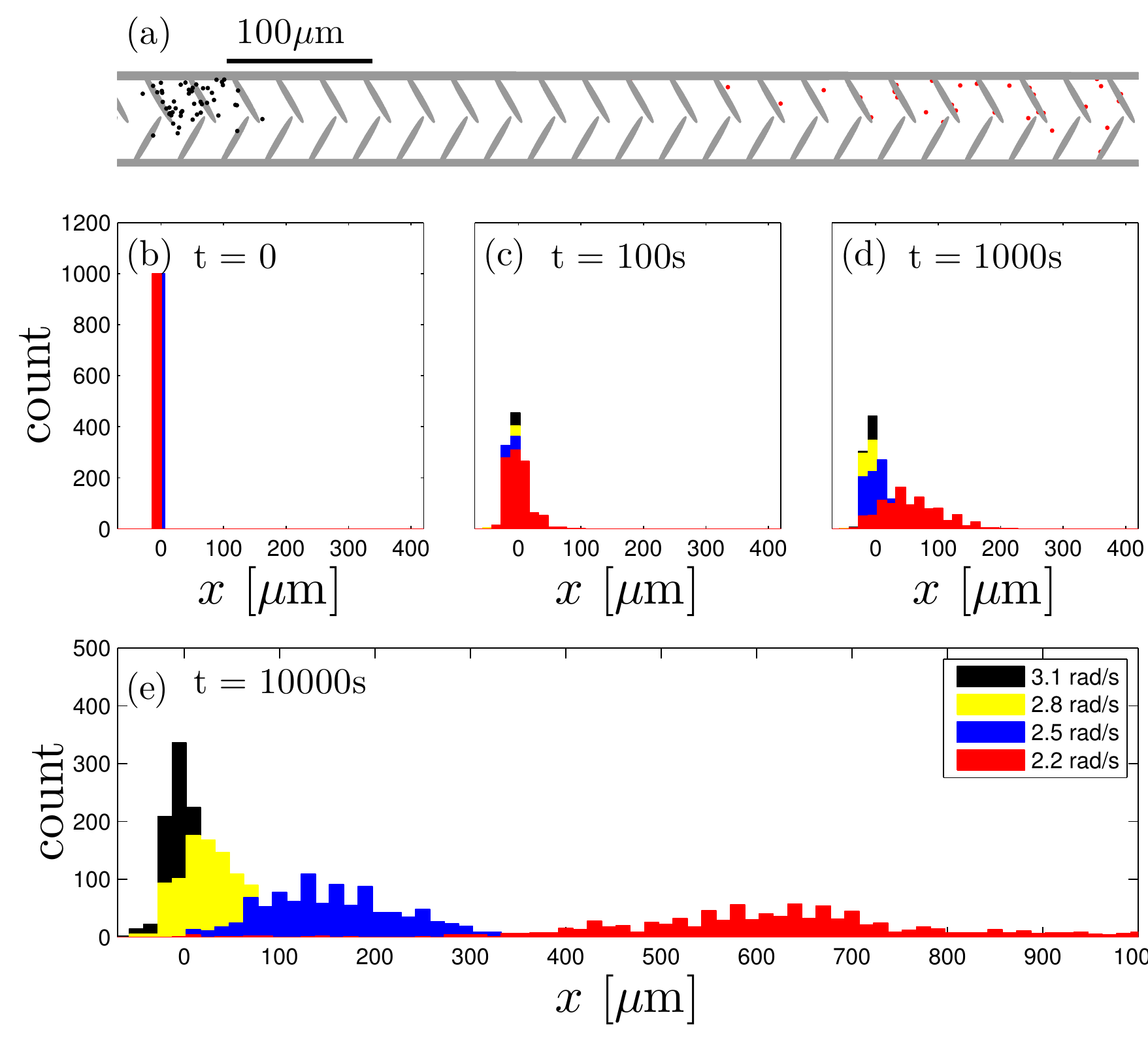}
\caption{Angular velocity based sorting in a microchannel. $(a)$ Patterned microchannel (grey areas) and position of levogyre $1000\,\mathrm{nm}$ microswimmers with $\Omega=2.2$ (black symbols) and $3.1\,\mathrm{rad/s}$ (red symbols) $1000\,\mathrm{s}$ after they have been released from position $x = 0\,\mathrm{\mu m}$ ($v=40\,\mathrm{\mu m/s}$, other parameters as in Tab.~\ref{tbl:table}). See also the supplementary movie 7. $(b-e)$ Histograms at various times of 1000 particles with $\Omega =2.2,\,2.5,\,2.8$ and $3.1\, \mathrm{\mu m/s}$ released at $t=0\,\mathrm{s}$ from $x = 0\,\mathrm{\mu m}$.
}
\label{fgr:figure5}
\end{figure}

\begin{figure}[h!]
\centering
\includegraphics[width=8.5cm]{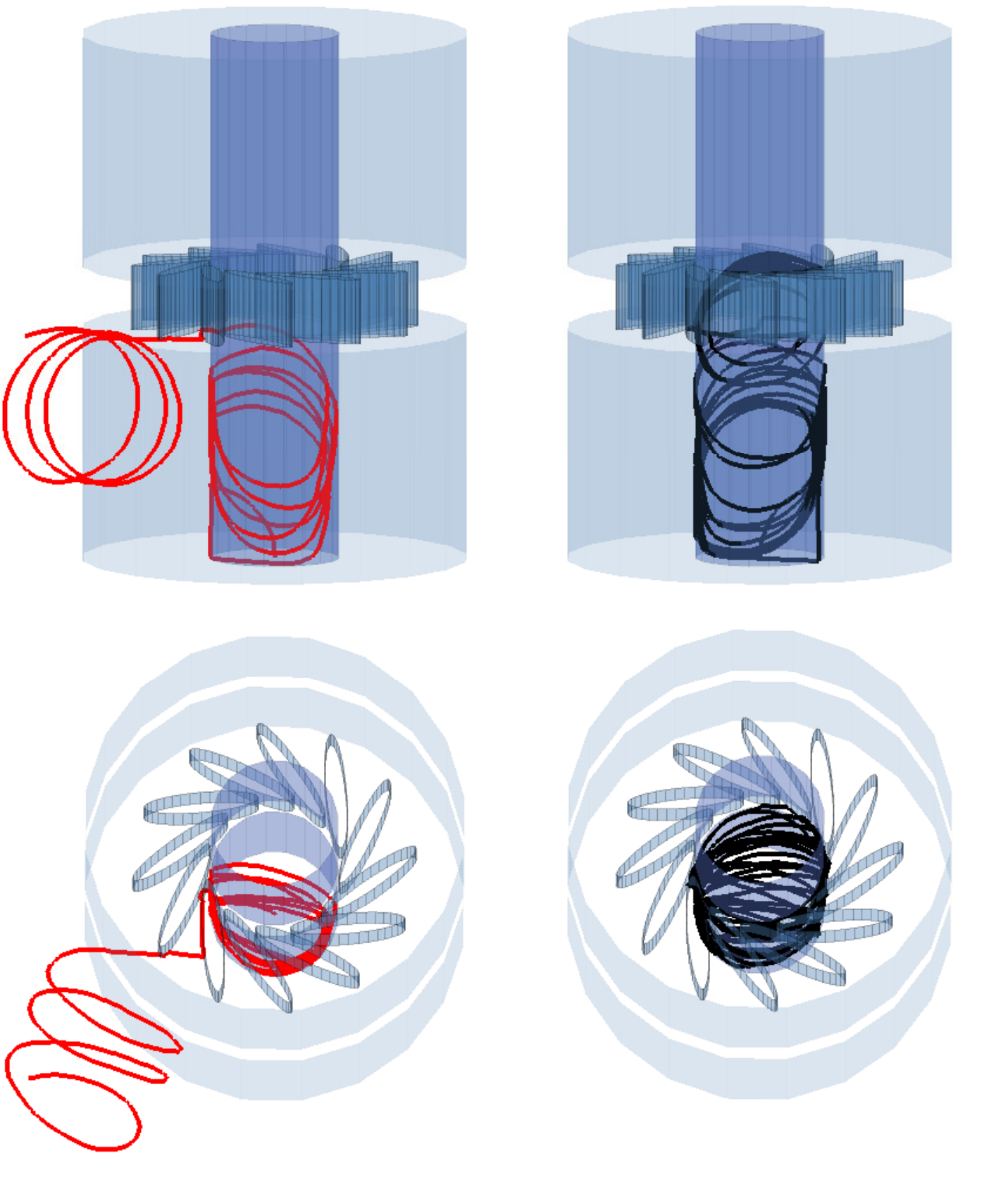}
\caption{Sorting of three-dimensional chiral microswimmers. The transparent structures represent a circular microchannel (inner diameter $10\,\mathrm{\mu m}$) with an engraved chiral flower. Views from the side (top) and from above (bottom) are presented. When levogyre microswimmers ($R = 1000\,\mathrm{nm}$, $v = 3.2$e+2$\,\mathrm{\mu m/s}$, $\Omega = +50 \,\mathrm{rad/s}$, red line on the left) are made to enter the channel from the lower end to break the reflection symmetry of the channel, they tend to exit the channel as soon as they reach the chiral flower, differently from dextrogyre microswimmers ($\Omega = -50 \,\mathrm{rad/s}$, black line on the right), which are prevented from escaping. The sorting efficiency is $Q=93\%$.
}
\label{fgr:figure7}
\end{figure}

The sorting mechanism discussed here for the case of two-dimensional chiral microswimmers performing circular motion can also be adapted to the case of three-dimensional chiral microswimmers performing helicoidal motion by using structures that are chiral in three dimensions. An example is demonstrated in Fig. \ref{fgr:figure7}. In a microchannel patterned with a chiral flower at a certain height, microswimmers moving in a given direction, e.g. because they are made to enter the tube from a given end, will behave differently depending on the chirality: levogyre microswimmers (red line on the left of Fig. \ref{fgr:figure7}) will tend to escape from the tube, while dextrogyre microswimmers (black line on the right of Fig. \ref{fgr:figure7}) will tend to remain in the tube. With the structure presented in Fig. \ref{fgr:figure7}, we obtained a sorting efficiency $Q=93\%$. It is also possible to extend to the sorting of three-dimensional microswimmers the approaches proposed in Figs. \ref{fgr:figure4} and \ref{fgr:figure5} by decorating the inside walls of the microchannel with chiral patterns. A similar approach has been implemented to sort passive chiral microswimmers making use of hydrodynamic interactions \cite{Aristov2013}; however, the advantage of employing active chiral particles, instead of relying on hydrodynamic interactions, is that it is possible to achieve much higher sorting efficiency in shorter times and distances.

\section{Conclusions and Outlook}

We have studied the sorting of chiral microswimmers on the basis of the chirality, linear velocity and angular velocity of their motion. This is obtained by using an environment with chiral patterns, which can be readily fabricated by standard microfabrication techniques. The main advantage lays in the simplicity of these methods, which only rely on static environmental structures and on the microswimmers' own self-propulsion, without requiring moving parts, fluid flows or external forces. These techniques can be exploited to separate naturally occurring chiral active particles, e.g., bacteria and spermatozoa, and also to separate chiral passive particles, e.g., chiral molecules, by coupling them to some chiral motors.

\subsection*{Acknowledgements}

We gratefully acknowledge Clemens Bechinger, Ivo Buttinoni and Felix K\"ummel for inspiring discussions and to O\v{g}uzhan Y\"{u}cel and Stefan Ristevski for help with the preparation of the plots. This work has been partially financially supported by the Scientific and Technological Research Council of Turkey (TUBITAK) under Grants 111T758 and 112T235, and Marie Curie Career Integration Grant (MC-CIG) under Grant PCIG11 GA-2012-321726.

\footnotesize{
\bibliography{rsc} 
\bibliographystyle{rsc} 
}

\section{Supplementary Materials}

\subsection{Supplementary Movie 1}
Trajectories of chiral levogyre (red) and dextrogyre (black) microswimmers with $R = 1\,\mathrm{\mu m}$ (see Tab.~\ref{tbl:table} for the other parameters), corresponding to Fig.~\ref{fgr:figure1}c.

\subsection{Supplementary Movie 2}
Trajectories of chiral levogyre (red) and dextrogyre (black) microswimmers with $R = 0.50\,\mathrm{\mu m}$ (see Tab.~\ref{tbl:table} for the other parameters), corresponding to Fig.~\ref{fgr:figure1}d.

\subsection{Supplementary Movie 3}
Trajectories of chiral levogyre (red) and dextrogyre (black) microswimmers with $R = 0.25\,\mathrm{\mu m}$ (see Tab.~\ref{tbl:table} for the other parameters), corresponding to Fig.~\ref{fgr:figure1}e.

\subsection{Supplementary Movie 4}
Trajectories of a levogyre (red line) and dextrogyre (black line) microswimmer with $R=1.00\,\mathrm{\mu m}$ (see Tab.~\ref{tbl:table} for the other parameters) in a chiral flower of ellipses (shaded areas), corresponding to Fig.~\ref{fgr:figure2}a.

\subsection{Supplementary Movie 5}
Separation of levogyre (red symbols) and dextrogyre (black symbol) microswimmers with $R=1.00\,\mathrm{\mu m}$ (see Tab.~\ref{tbl:table} for the other parameters) in a chiral patterned environment, corresponding to Fig.~\ref{fgr:figure3}.

\subsection{Supplementary Movie 6}
Separation of $1.00\,\mathrm{\mu m}$ microswimmers with $v=40$ (black symbols) and $70\,\mathrm{\mu m/s}$ (red symbols) $1000\,\mathrm{s}$ (other parameters as in Tab.~\ref{tbl:table}) in a patterned microchannel, corresponding to Fig.~\ref{fgr:figure4}a.
 
\subsection{Supplementary Movie 7}
Separation of $1.00\,\mathrm{\mu m}$ microswimmers with $\Omega=2.2$ (black symbols) and $3.1\,\mathrm{rad/s}$ (red symbols) $1000\,\mathrm{s}$ after they have been released from position $x = 0\,\mathrm{\mu m}$ ($v=40\,\mathrm{\mu m/s}$, other parameters as in Tab.~\ref{tbl:table}) in a patterned microchannel, corresponding to Fig.~\ref{fgr:figure5}a.

\end{document}